**Optical Metamagnetism and Negative Index Metamaterials**

Uday K. Chettiar, Alexander V. Kildishev, Wenshan Cai, Hsiao-Kuan Yuan, Vladimir P. Drachev, Vladimir M. Shalaev*

**Abstract**

A new class of artificially structured materials called *metamaterials* makes it possible to achieve electromagnetic properties that do not exist in nature. In this paper we review the recent progress made in the area of optical metamaterials. It was predicted that nanostructured metamaterials could provide us with artificial magnetic response and negative refractive index at optical frequencies. To date, *optical metamagnetics* have been already fabricated to demonstrate artificial magnetic response in the infrared and across the entire visible spectrum, while metamaterials showing negative refractive index, also called *negative index materials* (NIM), have been demonstrated in the infrared and at the border with the visible spectral range. Here we report the results of a sample that displays NIM behavior for red light at a wavelength of 710 nm. This is the shortest wavelength so far at which NIM behavior has been observed for light (excluding a device built on surface plasmon polaritons). We also discuss the fabrication challenges and the impact of fabrication limitations, specifically surface roughness of the fabricated structures on the optical properties of the metamaterials.

**Introduction**

The refractive index ( $n = n' + in''$ ) is one of the most important optical characteristics of any material. Conventionally, the refractive index ( $n'$ ) is assumed to be positive, but a negative refractive index ( $n' < 0$ ) does not violate any fundamental laws of physics. Negative index metamaterials (NIMs), with $n' < 0$, have some remarkable properties that make them candidates for a number of potential applications like super resolution.[1] In a landmark paper in 1968, Veselago showed that a material with simultaneous negative permeability ($\mu$) and permittivity ($\varepsilon$) has a negative refractive index [2]. However, this is actually strong (sufficient) condition for a negative index. The general condition for negative index is given by $\varepsilon'\mu'' + \mu'\varepsilon'' < 0$. The latter strictly implies that there cannot be $n' < 0$ in a passive metamaterial with $\mu = 1 + 0i$. Consequently, a magnetic response is essential in a NIM. Unfortunately, there is no magnetic response in nature at optical frequencies; hence to obtain an optical NIM we need to achieve an artificial magnetic response as a prerequisite. Typically, a NIM is an artificially engineered metal-dielectric composite that exhibits $n' < 0$ within a particular wavelength range.

Following from the above discussion, two types of NIMs can be introduced. A double negative NIM (DN-NIM) is a metamaterial with simultaneously negative real parts of effective permeability and permittivity ($\mu' < 0$ and $\varepsilon' < 0$). A single negative NIM (SN-NIM) has a negative refractive index with either $\mu' < 0$ or $\varepsilon' < 0$ (but not both). At optical wavelengths, obtaining $\varepsilon' < 0$ is easy compared with obtaining $\mu' < 0$, since noble metals naturally have a negative $\varepsilon'$ above the plasma wavelength.

The ratio $-n'/n''$ is often taken as a figure of merit (FOM) of NIMs since low-loss NIMs are desired. The FOM can be rewritten as $-\left(|\mu|\varepsilon' + |\varepsilon|\mu'\right)/\left(|\mu|\varepsilon'' + |\varepsilon|\mu''\right)$, indicating that a DN-NIM with $\mu' < 0$ and $\varepsilon' < 0$ is better than a SN-NIM with the same $n' < 0$ but with $\mu' > 0$. Since a DN-NIM will have a lower $n''$ when compared to a SN-NIM with the same value of $n'$. In addition, DN-NIMs can provide better impedance matching as compared to SN-NIMs.[3]

In this paper we review some of our and other groups work related to the demonstration of artificial magnetism and a negative index at optical frequencies. We specifically focus on our recent work on optical metamagnetics fabricated for the whole visible spectrum from red to blue. We also review our results regarding the demonstration of DN-NIM behavior close to 810 nm, which is the shortest wavelength at which DN-NIM behavior has been observed thus far. In addition we report SN-NIM behavior at 710 nm, which is the shortest wavelength at which a negative index has been shown. It should be noted that our work deals with negative indices for light. Negative index behavior has also been demonstrated at shorter wavelengths, but for surface plasmon polaritons (SPPs) rather than for light itself.[4] We also note that SPP properties of the two materials that form the interface on which the SPPs propagate and consequently it is hard to define a refractive index with its usual meaning.

**Metamagnetics**

As mentioned before, we need a negative permeability and permittivity for obtaining a negative refractive index. Obtaining negative permittivity at optical frequencies is straightforward since plasmonic materials exhibit a negative permittivity at frequencies below the plasma frequency, and common plasmonic materials like gold and silver have plasma frequencies just above the visible range. But unlike permittivity, the permeability of all naturally occurring materials is essentially equal to one at optical frequencies. Hence, as a first step to achieving negative refractive index, we need to come up with materials that exhibit not only a non-unity value of permeability, but also, ideally, a negative permeability. But it should be noted that even a non-unity value of permeability is of immense importance and can find useful applications in such diverse areas as subwavelength waveguides and antennas [5, 6], filters [7] and electromagnetic cloaking devices [8].

Various structures like bi-helixes,[9] split-ring resonators (SRR) [10] and pairs of metal nanorods [11-13] have been proposed for *metamagnetics* (metamaterials with artificial magnetic response). The bi-helix media has also been experimentally realized. [9] The SRR geometry has been extensively used to demonstrate artificial magnetism from the microwave frequency range to near infrared wavelengths up to 800 nm[14, 15]. However, the feasibility of the SRR geometry in the optical range is restricted due to inherent saturation effect that limits its magnetic response [16]. Therefore, in our work we have employed a nanostrip pair design that has been used to demonstrate artificial magnetic responses over the entire visible spectrum [17, 18].

Let's consider first the experimental observation of a negative permeability in the visible range [17, 18]. Figure 1 shows a schematic cross section of a unit cell of the negative permeability structure, which is a periodic array of pairs of this silver strips, along with a SEM image of the fabricated structure. Although the schematic shows the cross section to be rectangular in shape, the physical samples have a trapezoidal cross section due to limitations in fabrication. The structure was fabricated using electron beam lithography, and the results were reported in [17]. For the study, two samples (denoted as Samples A and B) with slightly different geometries and different silver surface roughnesses were fabricated. A negative effective permeability was retrieved using numerical simulations;[17] the results are in good agreement with the transmission and reflection spectra obtained from optical measurements for each sample. The value of $\mu'$ is about -1 in Sample A and about -1.7 in Sample B at the wavelengths of 770 nm and 725 nm, respectively.

To fabricate our negative permeability samples, electron beam lithography techniques have been used. First, the geometry of the periodic thin silver strips was defined by use of an electron beam writer on a glass substrate initially coated with a 15-nm film of indium-tin-oxide (ITO). Then, electron beam evaporation was applied to produce a stack of lamellar films. The deposition process was optimized by fabricating a number of samples under different conditions including the rate of deposition, which was varied from 80 Å/s to 0.5 Å/s. This was done to characterize the effect of deposition rate on the quality and optical properties of the final structure. Finally, a lift-off process was performed to obtain the desired silver strips.

Measurements of the root-mean square (RMS) value of surface roughness using a Veeco DI3100 atomic-force microscope (AFM) indicated that a slower deposition rate of silver (0.5 Å/s, Sample B) resulted in lower surface roughness than a faster deposition rate (~2Å/s, Sample A).  Note that for a *typical* deposition procedure, the rate cannot be set lower than 2 Å/s, since the deposition of silver heats up the resist material and makes the lift-off process impossible. To cool down the resist, we performed a novel 4-step deposition process with 10-minute pauses between each deposition step. This new procedure allows the use of a slower deposition rate of about 0.5 Å/s in order to obtain a lower surface roughness while ensuring successful lift-off, providing an overall better quality sample. The vertical layer structure of the films from the ITO-coated glass was: Sample A, 10 nm alumina, 30 nm silver, 40 nm alumina, 30 nm silver, 10 nm alumina; Sample B, 10 nm alumina, 35 nm silver, 40 nm alumina, 35 nm silver, 10 nm alumina. A representative SEM image of the fabricated structure of Sample A is shown in Figure 1(b), while Sample B is shown in Figure 1(c). The images were taken using a Hitachi S-4800.

   To test the fabricated samples, we measured the transmission and reflection spectra of the samples with an ultra-stable tungsten lamp (B&W TEK BPS100). The spectral range of the lamp covers the entire visible and near-infrared optical band. A Glan Taylor prism was placed at the output of the broadband lamp to select light with desired linear polarization. The signal transmitted (or reflected) from the sample was introduced into a spectrograph (Acton SpectraPro 300i) and eventually collected by a liquid-nitrogen cooled CCD-array detector. The transmission and reflection spectra were normalized to a bare substrate and a calibrated silver mirror, respectively. In the transverse electric (TE) regime the electric field of the incident light was linearly polarized parallel to the length of silver strips, while in the transverse magnetic (TM) mode the electric field was rotated 90 degrees relative to TE case. For example, Figures 2(a) and 2(b) show transmission and reflection spectra obtained from the optical measurements of Sample A and Sample B for TM polarizations at normal incidence along with the simulated results. Figures 2(c) and 2(d) show the real parts of the retrieved effective permeability and permittivity for Samples A and B, respectively. Sample A and B show negative permeabilities of –1 and –1.7 at 770 nm and 725 nm respectively. In the TE polarization, both samples act like a "diluted metal" film without any resonant response.[17]

It has been shown that roughness in the nanostrip structures along with finite size effects result in a decreased negative permeability due to additional losses caused by increased electron scattering.[19] In this case these effects result in a decreased negative permeability relative to the ideal case by factor 7.8 for Sample A and only by a factor 2.4 for Sample B, which has better surface quality.[17] These results are consistent with the statistical analysis of surface roughness obtained from AFM scans. The example cross sections of the AFM images shown in Figure 3(a) for Sample A and in Figure 3(b) for Sample B indicate substantial differences in the surface roughness of the samples. To extract quantitative characteristics the root mean square (RMS) roughness analysis was performed using seven 60x60 nm areas. For each sample the areas were selected at the tops of randomly chosen strips. Sample A is characterized by larger variations in RMS roughness across seven areas ranging from 2 to 6 nm (versus 1.5 – 2.5 nm variations for Sample B) with an RMS roughness value over the total area of about 3.7 nm (versus 2 nm for Sample B).

In summary, there are substantial deviations from the ideal, smooth silver strip surface and these deviations result in enhanced absorption at the resonances which were observed experimentally in the resonant plasmonic elements of the nano-structured samples. These non-idealities and the resulting deviations from the ideal properties can be reduced by improving the fabrication methods to minimize the imperfections. It is observed that using a slower deposition decreases the surface roughness and gives improved magnetic response.

### *Extension of metamagnetism to the entire visible spectrum*

The results mentioned above were later extended to the entire visible spectrum by fabricating a family of nanostrip pairs with a range of widths to provide magnetic resonance across the visible spectrum.[18] Figure 4 shows the optical microscope images of the array of samples for TE and TM polarization in reflection and transmission mode. In the resonant TM polarization, we observe distinct colors in the different samples both in transmission mode and reflection mode, indicating a unique resonant frequency in each sample. On the other hand, for the non-resonant TE polarization, the colors are the same for all the samples. In this case the samples act as "diluted metal" films with more reflection and less transmission at longer wavelengths.

The metamagnetic samples were optically characterized to obtain their spectral response. Figure 5 shows the transmission and reflections of the six samples in the TM polarization. We see strong resonant behavior in all the samples for wavelengths ranging from 491 nm to 754 nm, covering the majority of the visible spectrum. The position of the magnetic resonance moves towards the shorter wavelengths as the width of the nanostrips pairs decreases. Figure 5(c) shows the real part of permeability for the six samples as a function of the resonance wavelengths. The permeability ranges from -1.6 at 750 nm to 0.5 at 500 nm.

**Negative Index Metamaterials**

In the previous section we saw that an array of metal nanostrip pairs can exhibit a negative permeability. A modification of this structure, sometimes known as the double grating structure or the fishnet structure, is widely used to demonstrate negative indices at optical frequencies. This structure has been used to demonstrate NIM behavior at 780 nm[20], 1.5 µm[21] and 2 µm[22]. However in all these cases, only SN-NIM behavior was observed. The first DN-NIMs in the optical range were demonstrated at 1.4 µm with a FOM of about three[21] and 1.8 µm with a FOM above 1[23]. The shortest wavelength at which DN-NIM behavior was observed is about 810 nm with a FOM of 1.3[3]. The same sample also demonstrated SN-NIM behavior, for a different polarization close to 770 nm with a FOM of 0.7. Here we review the results from this sample as well as report new results obtained from a sample showing NIM behavior at the shortest wavelength to date. This recent sample exhibits SN-NIM behavior at 710 nm with a FOM of 0.5.

The structure was fabricated using a procedure similar to that described above for fabricating the magnetic samples. In the fabrication process, a slow deposition rate was used in order to minimize the roughness of the metal surface. The structures were characterized by measuring the spectral response. Figures 6(a) shows the SEM image of the sample. Figure 6(b) shows the simulated and experimental spectra of the sample. The experimental and simulated spectra shown in Figure 6b demonstrate a good match over a broad range of measured wavelengths (from 500 to 950 nm) and include sharp resonance features around 600 nm and 800 nm. Simulations provide comprehensive proof on the origin of these resonances.[3] Figure 7(a) shows a schematic of a unit cell of the fishnet structure. The region marked with the red hashing is the area which is responsible for the resonances. Figure 7(b) and (c) depict the field maps of the highlighted region at 625 nm and 815 nm respectively. The absorbance spectrum shows strong resonant behavior at these two wavelengths. In the field maps the color map represents the magnetic field which is normalized with respect to the incident magnetic field. The arrows represent the electric displacement. In Figure 7(b) we can see that the electric displacement in both the top and bottom metal strips are aligned together resulting in strong electric dipole response which implies a resonance that is electrical in nature. In Figure 7(c) the electric displacement in the two metal strips are oppositely directed resulting in a magnetic dipole moment which is evident from the strong magnetic field between the two strips. Hence, this resonance is termed as a magnetic resonance. It is worth noting that the magnetic field in the dielectric space is negative with respect to the incident field resulting in a negative permeability as we shall see ahead.

The retrieved results for the sample are shown in Figure 8. The real part of the refractive index ($n'$) and the figure of merit (FOM = $-n'/n''$) are depicted in Figure 8(a) (For the sake of clarity FOM is set to zero when $n' > 0$). The best FOM of 1.3 is obtained at a wavelength of 813 nm, where $n'$ is about –1.0. The minimum value of the refractive index, $n' \approx -1.3$, is achieved at 820 nm, but with a lower FOM of 0.9. As indicated in Figure 8(b), $\mu'$ is negative between 799 and 818 nm. This band is the DN-NIM regime. In this wavelength range, $\varepsilon'$ varies within the range of –1.2 ±0.1. The strongest magnetic response ($\mu' \approx -0.7$) is obtained at a wavelength of 813 nm, where $\varepsilon' \approx -1.1$. The same sample also demonstrates a SN-NIM response at the orthogonal polarization. In the orthogonal polarization the structure shows a maximum FOM of 0.7 at 772 nm with $n' \approx -1.0$.[3]

We also report an SN-NIM behavior at the shortest wavelength to date. The SEM image of the sample is shown in Figure 9(a). The simulated and experimental spectra are shown in Figure 9(b). We can clearly see the strong resonant response around 710 nm. There is also good agreement between the experimental and simulated results. The simulated results were used to retrieve the effective refractive index for the sample and are shown in Figure 9(c). As we can see the structure shows a maximum FOM of 0.5 at 710 nm with $n' \approx -0.6$. This is the shortest wavelength at which negative index has been demonstrated for light.

**Conclusions**

Metamaterials have the potential of adding another dimension to the set of existing materials by providing us with novel properties that do not exist in nature. Optical metamagnetics and negative index metamaterials are two such examples. We have seen that optical metamagnetics allow us to obtain magnetic response in the infrared and across the entire visible spectrum in spite of the fact that no natural material exhibits magnetic behavior at these frequencies. Negative refractive index has been demonstrated at the optical wavelengths using various geometries like paired nanorod arrays and fishnets. Of these, the fishnet geometry shows more promise, and it has been used to demonstrate a DN-NIM response at wavelengths as short as 813 nm. The sample showed a maximum figure of merit of 1.3 with $n' \approx -1.3$. In this paper, we have also reported the shortest wavelength at which NIM behavior has been observed. The sample displayed a SN-NIM response at 710 nm with a maximum FOM of 0.5 with $n' \approx -0.6$. We have also showed that the quality of the nanostructured metal in metamagnetics and NIMs can have substantial impact on their performance. By using optimized fabrication procedures, there is a significant room for improving the metamaterial properties. We observed that by utilizing a slower deposition rate the surface roughness of the structures can be reduced resulting in a much better response. As the field of metamaterials continues to grow and expand, we anticipate a variety of unique and exciting applications in optics based on this new class of materials.

## Acknowledgements

This work was supported by ARO-MURI award 50342-PH-MUR and NSF-PREM grant #DMR-0611430.

**Figure Captions**

1. Figure 1. (a) Schematic of a pair of nanostrips, (b) SEM image of sample A, (c) SEM image of sample B.

2. Figure 2.(a,b) Spectra of Sample A and Sample B respectively. Solid lines represent experimental data and dashed lines represent simulated data. T is transmittance and R is reflectance. (c,d) Real parts of retrieved effective permeability and permittivity for Sample A and Sample B respectively.

3. Figure 3. (a,b) Representative AFM cross section of Sample A and Sample B respectively.

4. Figure 4. Optical microscope image of the array of magnetic samples for two orthogonal polarizations. (a) Transmission mode with TM polarization, (b) Transmission mode with TE polarization, (c) Reflection mode with TM polarization, (d) Reflection mode with TE polarization.

5. Figure 5. (a,b) Transmittance and reflectance of the array of magnetic samples, (c) Real part of permeability vs. resonance wavelength for the six samples in the array of magnetic samples.

6. Figure 6. (a) FE-SEM image of the DN-NIM sample, (b) Experimental and simulated spectra of the DN-NIM sample. T is transmittance, R is reflectance and A is absorbance with solid lines and indices e representing experimental data and dashed lines and indices s representing simulated data.

7. Figure 7. (a) Schematic figure of a unit cell of the fishnet structure along with the incident polarization . The region highlighted with the red hashing is the region of resonance. (b,c) Field maps of the region highlighted in (a) at a wavelength of 625 nm and 815 nm respectively. The color represents the magnetic field normalized with respect to the incident and the arrows represent the electric displacement.

8. Figure 8. (a) Real part of refractive index and FOM for the DN-NIM sample, (b) Real part of permittivity and permeability for the DN-NIM sample.

9. Figure 9. (a) FE-SEM image of SN-NIM sample, (b) Experimental and simulated spectra for the SN-NIM sample. T is transmittance, R is reflectance and A is absorbance with solid lines and indices e representing experimental data and dashed lines and indices s representing simulated data.    (c) Real part of refractive index and FOM for the SN-NIM sample.

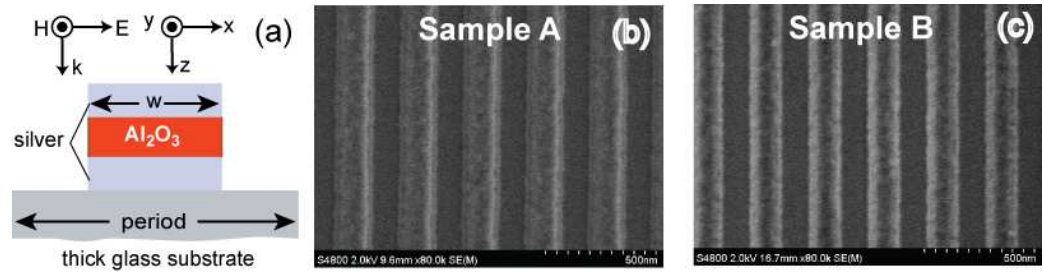

Figure 1. (a) Schematic of a pair of nanostrips, (b) SEM image of sample A, (c) SEM image of sample B .

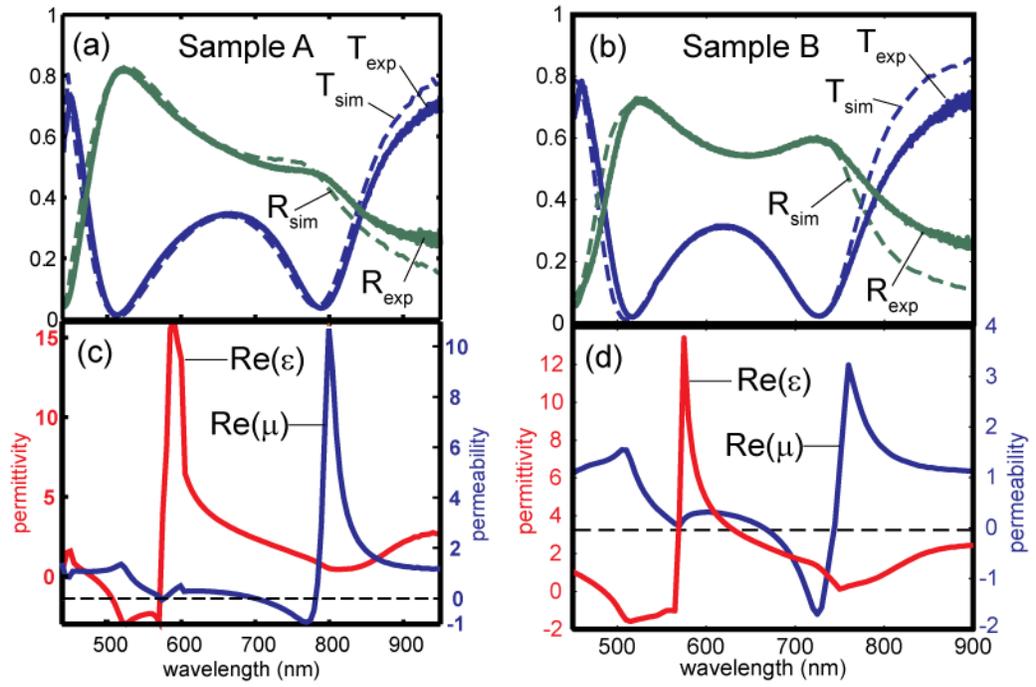

Figure 2.(a,b) Spectra of Sample A and Sample B respectively. Solid lines represent experimental data and dashed lines represent simulated data. T is transmittance and R is reflectance. (c,d) Real parts of retrieved effective permeability and permittivity for Sample A and Sample B respectively.

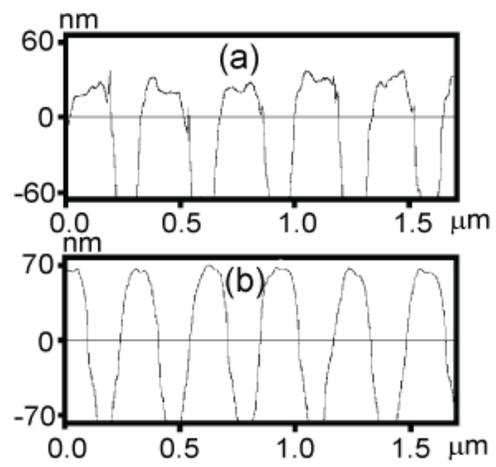

Figure 3. (a,b) Representative AFM cross section of Sample A and Sample B respectively.

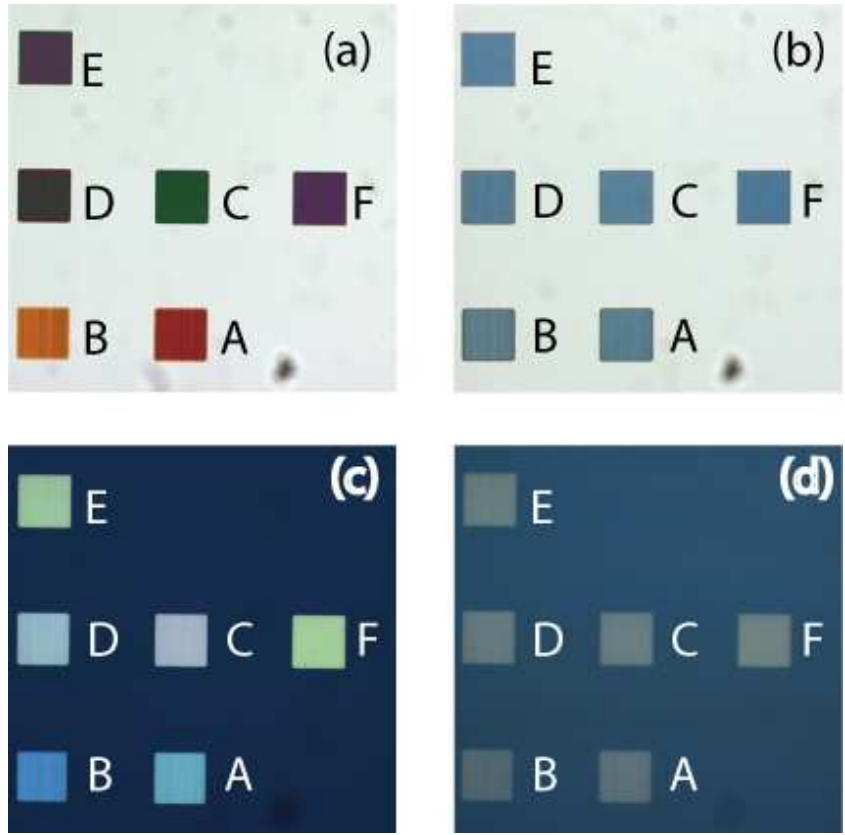

Figure 4. Optical microscope image of the array of magnetic samples for two orthogonal polarizations. (a) Transmission mode with TM polarization, (b) Transmission mode with TE polarization, (c) Reflection mode with TM polarization, (d) Reflection mode with TE polarization.

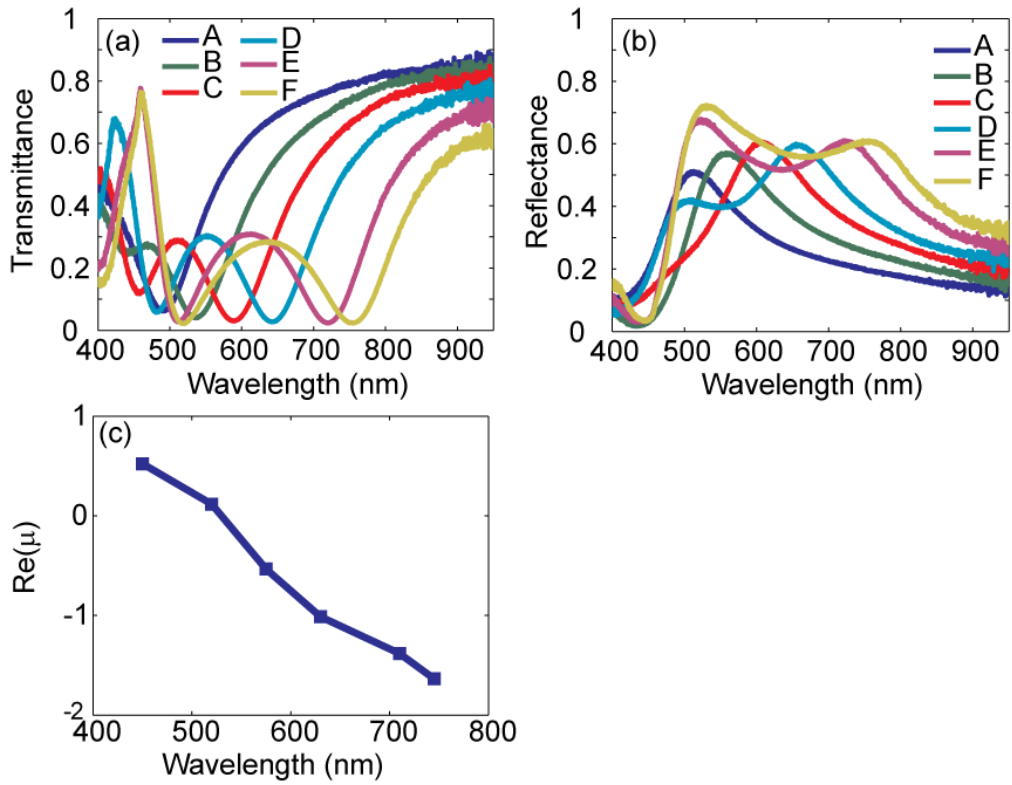

Figure 5. (a,b) Transmittance and reflectance of the array of magnetic samples, (c) Real part of permeability vs. resonance wavelength for the six samples in the array of magnetic samples.

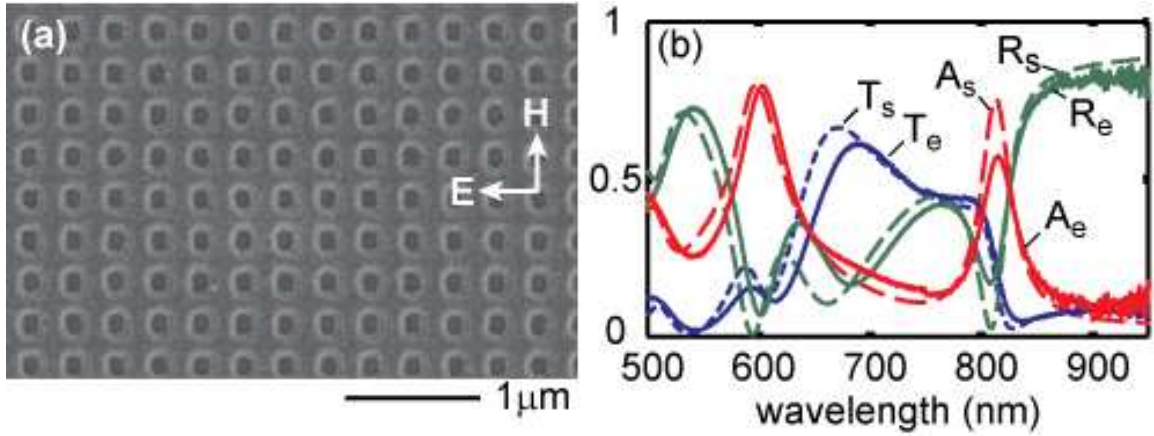

Figure 6. (a) FE-SEM image of the DN-NIM sample, (b) Experimental and simulated spectra of the DN-NIM sample. T is transmittance, R is reflectance and A is absorbance with solid lines and indices e representing experimental data and dashed lines and indices s representing simulated data.

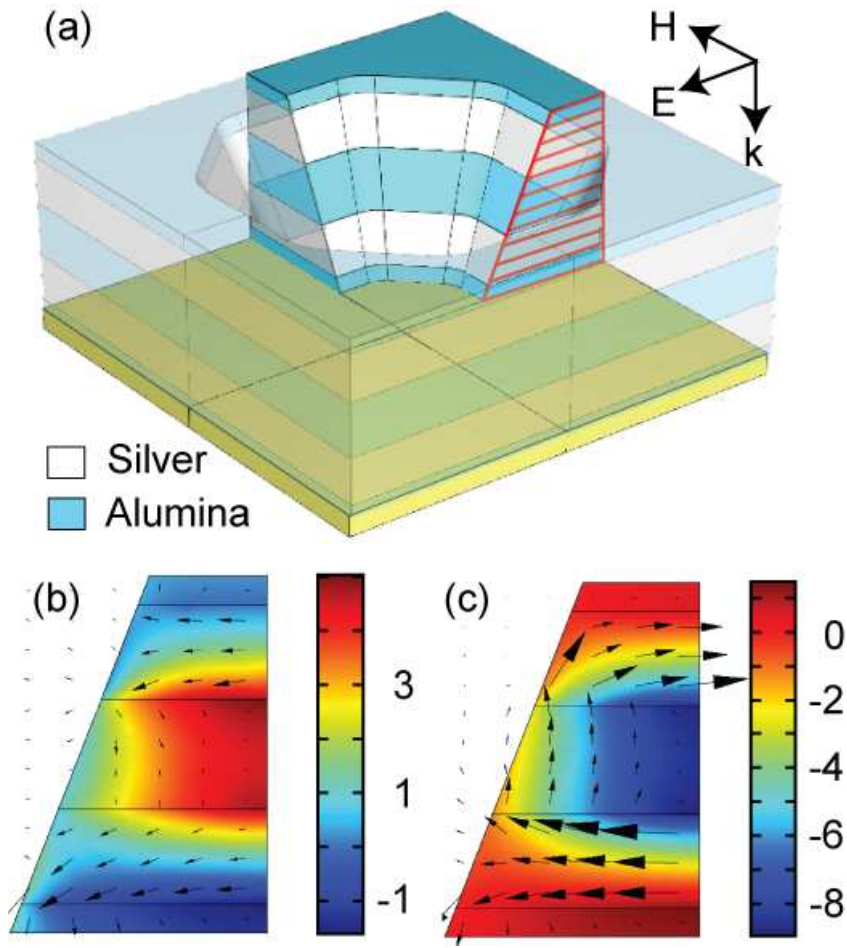

Figure 7. (a) Schematic figure of a unit cell of the fishnet structure along with the incident polarization . The region highlighted with the red hashing is the region of resonance. (b,c) Field maps of the region highlighted in (a) at a wavelength of 625 nm and 815 nm respectively. The color represents the magnetic field normalized with respect to the incident and the arrows represent the electric displacement.

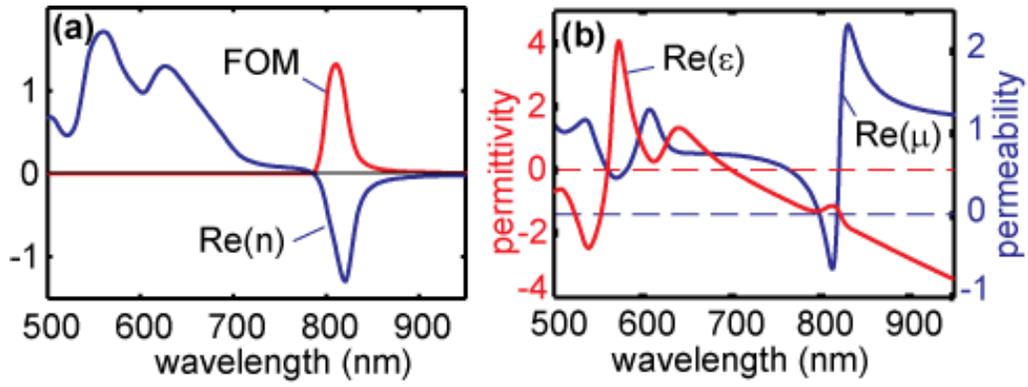

Figure 8. (a) Real part of refractive index and FOM for the DN-NIM sample, (b) Real part of permittivity and permeability for the DN-NIM sample.

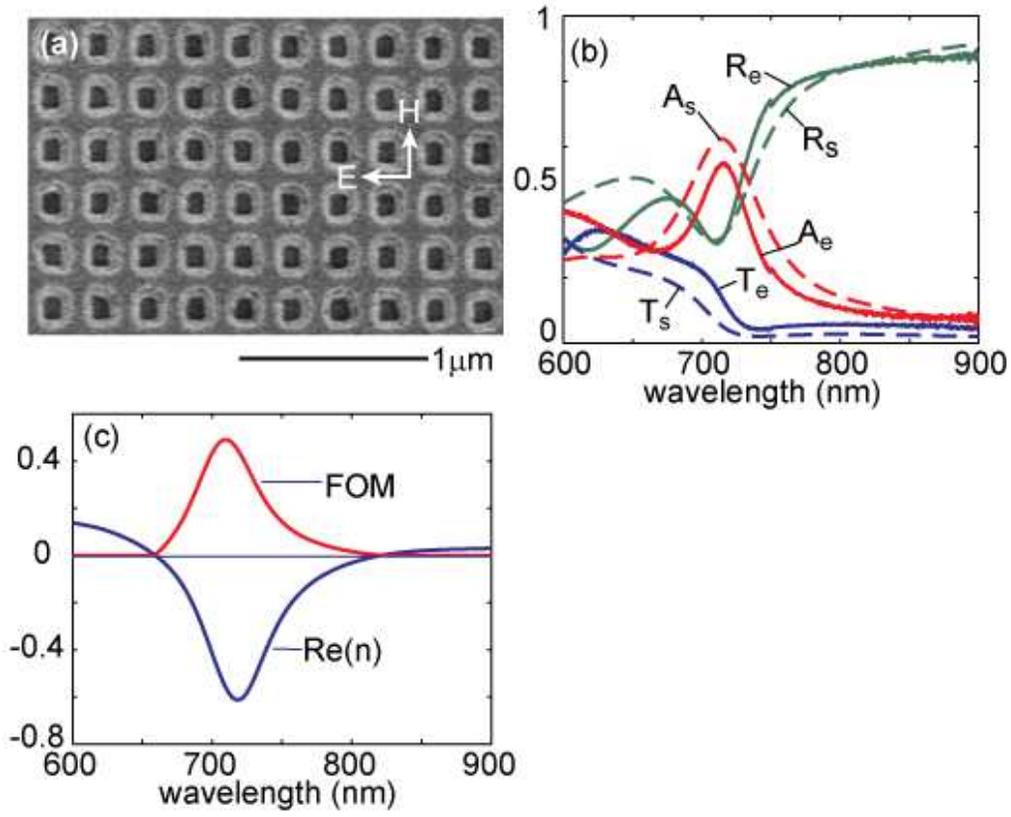

Figure 9. (a) FE-SEM image of SN-NIM sample, (b) Experimental and simulated spectra for the SN-NIM sample. T is transmittance, R is reflectance and A is absorbance with solid lines and indices e representing experimental data and dashed lines and indices s representing simulated data.   (c) Real part of refractive index and FOM for the SN-NIM sample.